# Brain volume is a better biomarker of outcomes in ischemic stroke compared to brain atrophy

**Kenda Alhadid[1], Robert W. Regenhardt[1], Natalia S. Rost[1], Markus D. Schirmer[1]***

[1]Department of Neurology, Massachusetts General Hospital, Harvard Medical School, Boston, MA, USA.

**\* Correspondence:**
Markus D. Schirmer
mschirmer1@mgh.harvard.edu

**Abstract**

**Objective:** To assess if brain volume at the time of ischemic stroke injury is a better biomarker of functional outcome than brain atrophy.

**Background:** Brain parenchymal fraction (BPF) has been used as a surrogate measure of global brain atrophy, and as a neuroimaging biomarker of brain reserve in studies evaluating clinical outcomes after brain injury. Brain volume itself is affected by natural aging, cardiovascular risk factors, and biological sex, amongst other factors. Recent works have shown that brain volume at the time of injury can influence functional outcomes, where larger brain volumes are associated with better outcomes.

**Methods:** Acute ischemic stroke cases at a single center between 2003 and 2011, with MR neuroimaging obtained within 48 hours from presentation were eligible. Functional outcomes represented by the modified Rankin Score (mRS) at 90 days post admission (mRS≤2 deemed a favorable outcome) were obtained via patient interview or per chart review. Deep learning enabled automated segmentation pipelines were used to calculate brain volume, intracranial volume (ICV), and BPF on the acute neuroimaging data. Patient outcomes were modeled through logistic regressions, and model comparison was conducted using the Bayes Information Criterion (BIC).

**Results:** 467 patients with arterial ischemic stroke were included in the analysis. Median age was 65.8 (interquartile range: 55.3-76.3) years, and 65.3% were male. In both models, age and a larger stroke lesion volume were associated with worse functional outcomes. Higher BPF and a larger brain volume were both associated with favorable functional outcomes, however, comparison of both models suggested that the brain volume model (BIC=501) explains the data better compared to the BPF model (BIC=511).

**Conclusions:** The extent of global brain atrophy (and its surrogate biomarker BPF) has been regarded as an important biomarker of post-stroke functional outcomes and resilience to acute injury. Here, we demonstrate that a higher global brain volume at the time of injury better explains favorable functional outcomes, which can be directly clinically assessed.

# 1 Introduction

With aging populations in the US and worldwide, and the increased incidence of stroke in younger patient populations, the prevalence of arterial ischemic stroke is increasing.(1) Understanding the determinants of post-stroke outcomes is of great clinical, societal, and economic importance. Determining the most relevant clinical and imaging biomarkers of functional outcomes is essential for developing targeted preventative and therapeutic approaches. Phenotypic information, such as age and lesion volume,(2–4) have been utilized to model post-stroke outcome, however, current models are insufficient to adequately explain clinically observed variations in outcomes.

Recently, neuroimaging studies revealed other important factors pertaining to clinical outcomes, such as white matter hyperintensity volume (WMHv).(5–7) Additionally, studies have demonstrated that brain volume, specifically cortical volume, is related to an individual's cognitive abilities and intelligence, even when corrected for age, sex and other collinearities.(8–12) Importantly, brain volume of stroke patients at the time of admission has been identified as an independent biomarker for functional post-stroke outcome.(13–15) Volumetric brain studies often normalize each patient's brain volume by their intracranial volume, also known as brain parenchymal fraction (BPF), which can serve as a surrogate measure of global brain atrophy in cross-sectional studies.(16) However, no consensus on the utility of non-normalized and normalized brain volume exists.

In this work, we utilize advances in deep-learning enabled segmentation algorithms to estimate brain volume and BPF in a cohort of 476 acute ischemic stroke patients based on their acute clinical neuroimaging data acquired in the emergency department or during hospital admission. Using multivariable logistic regression models of functional outcome, measured by the 90-day modified Rankin Scale (mRS) score, we compare the models including either BPF as a surrogate measure of brain atrophy or a volumetric measure of brain volume. We demonstrate that an individual's brain volume at the time of acute injury rather than a measure of brain atrophy, is a better marker for modeling functional outcome.

## 2 Materials and Methods

### 2.1 Standard protocol approvals, registration, and patient consent

The use of human patients in this study was approved by the local Institutional Review Board and informed written consent was obtained according to the Declaration of Helsinki from all participating patients or their surrogates at time of enrollment.

### 2.2 Study design, setting, and patient population

Patients over 18 years of age presenting to the emergency department at our hospital between 2003 and 2011 with signs and symptoms of acute ischemic stroke (AIS) were eligible for enrollment. In this analysis, we included subjects with (a) acute cerebral infarct lesions confirmed by diffusion weighted imaging (DWI) scans obtained within 48 hours of symptom onset and (b) T2 fluid-attenuated inversion recovery (T2-FLAIR) sequences available for volumetric analyses. All clinical variables including demographics and medical history were obtained on admission. Patients and/or their caregivers were interviewed in person or by telephone at 3 months after the acute clinical stroke presentation to assess functional outcome (mRS). If the patient could not be contacted, an mRS score was determined from review of clinical evaluations.

The standard AIS protocol included DWI (single-shot echo-planar imaging; one to five B0 volumes, 6 to 30 diffusion directions with b=1000 s/mm2, 1-3 averaged volumes) and T2 FLAIR imaging (TR 5000ms, minimum TE of 62 to 116ms, TI 2200ms, FOV 220-240mm). DWI data sets were assessed and corrected for motion and eddy current distortions.(17) Acute infarct volume was manually assessed on DWI (DWIv). A manual estimate of ICV was calculated on T1 sagittal sequences using a previously validated method.(18)

## 2.3 Automated brain and intracranial volume estimation

Estimation of brain and intracranial volume (ICV) was calculated in a standardized, automated process utilizing the available FLAIR imaging data. Each patient image first underwent N4 bias field correction,(19) followed by brain extraction using synthstrip.(20) The estimated brain mask was utilized in a secondary N4 bias field correction, after which the image underwent intensity normalization using a mean shift algorithm and normal appearing white matter was set to an intensity of 0.75. Subsequently, each image underwent thresholding at 0.375 to extract an estimate of total brain volume, given by combined gray and white matter volume, following our previously published approach.(14) ICV masks were estimated based on the segmentation results from synthseg,(21) utilizing the bias field corrected image as input. Each brain and ICV mask underwent manual quality control by visual inspection, and volumes were calculated by multiplying the number of voxels within the segmentation mask by the corresponding voxel volume. Figure 1 presents an overview of the full pipeline.

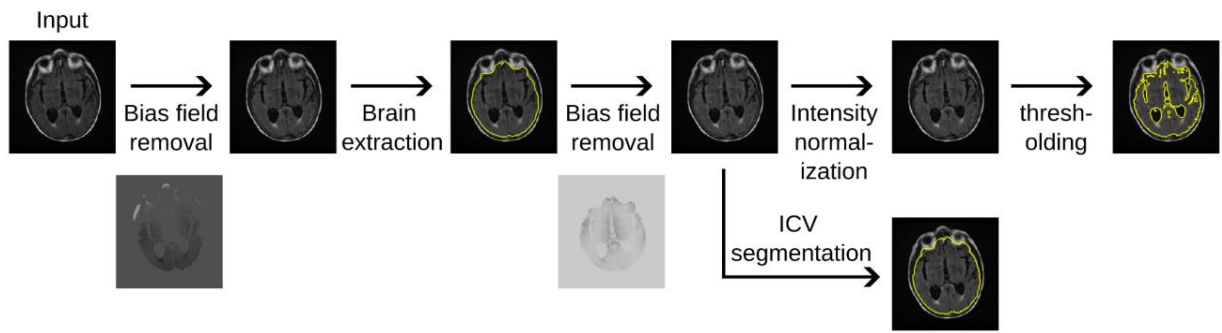

**Figure 1.** Processing pipeline for generating brain and ICV segmentations using clinical FLAIR sequences.

## 2.4 Statistical analysis and model description

Prior to analysis, each mask underwent manual quality control by visual inspection. For each patient, brain and intracranial volume was determined by multiplying the number of voxels by the corresponding voxel size. Automated and manual estimates of ICV were compared using a linear model without intercept, reporting the β coefficient.

We calculated BPF for each patient, given as the ratio of brain volume by intracranial volume, which was subsequently logit transformed. Age and brain volume were utilized in the model in units of decade and dm3, respectively, to avoid modeling issues due to scale. Patient outcome was encoded as functional independence (mRS≤2) and moderate to severe disability (mRS>2). Patient outcome was then modeled through logistic regressions, given as

$$mRS(>2) \sim Age + Sex + HTN + DM2 + Non-Smoker + VDWI + X,$$

where X was either BPF or brain volume, resulting in 2 models for comparison. Model comparison was conducted using the Bayes Information Criterion (BIC).

After model fit, we tested the model assumptions, i.e., linearity in the logit for continuous variables, absence of multicollinearity given by a variance inflation factor (VIF) lower than 2, and lack of strongly influential outliers. All statistical analyses were conducted using the computing environment R.(22) Significance was set at $p<0.05$.

## 2.5 Data availability statement

The authors agree to make the data, methods used in the analysis, and materials used to conduct the research available to any researcher for the express purpose of reproducing the results and with the explicit permission for data sharing by the local institutional review board.

## 3 Results

The clinical characteristics of the study cohort are described in Table 1. The cohort had a median age (interquartile range, IQR) of 65.8 (55.3, 76.3) years, 65.3% were male, 69.7% had a diagnosis of hypertension, and 24.8% of patients had a bad outcome with mRS>2. Manual and automated ICV estimates showed good agreement (coefficient ± standard error; $\beta = 0.944\pm0.002$).

**Table 1.** Characteristics of the cohort utilized in this study. (IQR: interquartile range; HTN: hypertensive; DM2: Diabetes Mellitus Type 2; BPF: brain parenchymal fraction)

| N | 476 |
|---|---|
| Age (years; median [IQR]) | 65.8 [55.3, 76.3] |
| Sex(% male) | 311 (65.3) |
| HTN (%) | 332 (69.7) |
| DM2 (%) | 96 (20.2) |
| Non-Smoker (%) | 286 (60.1) |
| Lesion Volume (cc; median [IQR]) | 2.2 [0.6, 12.7] |
| BPF (%; median [IQR]) | 0.81 [0.77, 0.83] |
| Brain volume (cc; median [IQR]) | 1306.9 [1190.9, 1413.9] |
| mRS (> 2; %) | 118 (24.8) |

The parameters of both outcome models are summarized in Table 2 and Figure 2. All assumptions of the logistic regression models were fulfilled. In both models, older patients and patients with larger stroke lesion volume had worse outcomes. Male sex was only found to be significant in the BPF outcome model, where male patients demonstrated better functional outcomes. Both higher BPF, i.e. less brain atrophy, and higher brain volume led to better functional outcomes.

Table 2. Summary of model parameter estimates. (HTN: hypertensive; DM2: Diabetes Mellitus Type 2; BPF: brain parenchymal fraction)

|  | BPF | p | Brain volume | p |
|---|---|---|---|---|
| **Intercept** | 2.95 | 0.366 | 1.78 | 0.266 |
| **Age** | **0.25** | **0.019** | **0.25** | **0.010** |
| **Sex (M)** | **-0.88** | **<0.001** | -0.36 | 0.194 |
| **HTN** | 0.20 | 0.506 | 0.08 | 0.792 |
| **DM2** | 0.54 | 0.056 | 0.47 | 0.105 |
| **Non-Smoker** | -0.08 | 0.737 | -0.02 | 0.919 |
| **log(Lesion Volume)** | **0.34** | **<0.001** | **0.35** | **<0.001** |
| **BPF** | **-7.32** | **0.038** | – |  |
| **Brain volume** | – | – | **-3.83** | **<0.001** |

Evaluating BIC for both models resulted in 511 and 501 for the model based on BPF and brain volume, respectively. The comparison of both models suggest that the brain volume model explains the observed data better than the corresponding base model, with ΔBIC = 10.

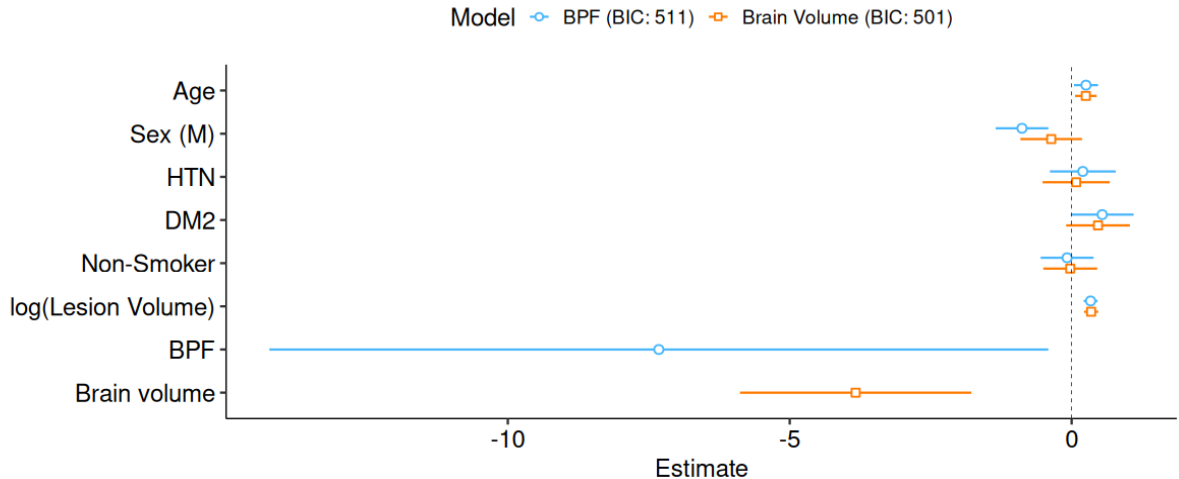

Figure 2. Graphical representation of parameter estimates including 95% confidence intervals. (BIC: Bayes Information Criterion; HTN: hypertensive; DM2: Diabetes Mellitus Type 2; BPF: brain parenchymal fraction).

## 4   Discussion

In this work, we highlight the significant role of brain volume and its association with functional outcomes after ischemic stroke. In a large cohort of AIS patients, we demonstrated that uncorrected brain volume at the time of injury is a better biomarker of stroke outcomes compared to brain atrophy. We derived brain volume and intracranial volume estimates automatically on clinical MRI sequences using a deep-learning enabled pipeline. This allowed the extraction of this important parameter from clinical imaging data obtained as standard of care for patients with acute stroke presentations. Our results indicate that a larger brain volume at the time of acute injury leads to better functional outcome. In two models comparing brain volume and BPF, we determined a ΔBIC = 10, providing strong evidence that the brain volume model outperforms the BPF model.(23)

The relationship between larger brain volume and higher cognitive abilities has been consistently reported,(8,24) with more recent work delineating the underlying microstructural architecture observed in larger brain volumes that could explain this benefit. It is put forth that larger cortices benefit from the increased processing power of a higher number of neurons, with concomitant lower neurite density and orientation dispersion maximizing network efficiency and reducing energy demand.(12) Importantly, other prior works have shown that total brain volume was a significant determinant of measured, and patient reported, functional outcomes after ischemic stroke.(14,15) This may further relate to the concepts of brain and effective reserve, which aims to quantify the brain's ability to compensate for negative effects, such as sudden vascular events.(25,26) Our data show that brain volume, without normalizing for intracranial volume, was a better determinator of functional outcomes post stroke.

In our model evaluating the relationship between BPF and functional outcomes at 90 days, male sex is significantly associated with a favorable functional outcome, and this is in line with the prior literature showing that females are typically older in age at time of stroke and endure worse functional outcomes post stroke.(27) However, sex becomes non-significant after including brain volume at time of injury. This is likely explained by brain volume differences in the context of known anthropometric differences between males and females, which may account for the majority of sex-specific variation in the current data. Future large cohort studies are needed to further disentangle sex-specific differences in patient outcomes.

There were limitations to our study. Due to time and resource limitations during an AIS clinical presentation, only a limited number of axial slices were obtained during MRI acquisition. These clinical scans lack isotropic resolution and experience partial volume effects. However, the imaging data used reflect the imaging data available during standard-of-care of these patients, supporting the findings are generalizable and immediately translatable. Identifying biomarkers represents an important aspect in improving patient outcomes. Furthermore, there is the potential of brain swelling due to the acute stroke lesion, which can artificially inflate brain volume during this time point. Ad-hoc analysis of the correlation between brain volume and log-transformed lesion volume, however, shows only low correlation between both variables (r=0.1; p=0.035), which agrees with prior literature.(14) Finally, treatment of stroke patients might modify their outcome. Granular treatment details were not available in our cohort. Considering that imaging occurred shortly after admission, however, it is unlikely that it would influence brain volume directly. Future large-scale studies with treatment information available are needed for further investigation.

Strengths of our study include the utilization of a large hospital-based cohort with clinical neuroimaging data available in the emergency department. Importantly, employing state-of-the-art clinical neuroimaging analysis methodologies enabled us to delve deeper into the associations with post-stroke outcomes at the time of hospital admission, identifying neuroimaging biomarkers that can readily be assessed in the clinical setting. To the best of our knowledge, this study represents the first investigation into the benefit of utilizing non-normalized volumetric estimates of brain volume over measures of brain atrophy.

## 5 Conclusion

Our study provides strong evidence in a large cohort of stroke patients that brain volume at the time of injury is a better determinant of functional post stroke outcomes compared to brain atrophy. Importantly, the presented analysis pipeline was based on clinical MRI, offering the opportunity for immediate translation to determine the biomarker of brain volume from routinely acquired clinical neuroimaging. This opens new avenues for expanding our existing knowledge on risks and outcomes in stroke populations.

## 6 Conflict of Interest

*The authors declare that the research was conducted in the absence of any commercial or financial relationships that could be construed as a potential conflict of interest.*

## 7 Funding

Research reported in this publication was supported by the National Institute of Aging of the National Institutes of Health under award number R21AG083559. NSR is supported by NINDS U19NS115388. RWR serves on a DSMB for a trial sponsored by Rapid Medical, serves as site PI for studies sponsored by Microvention and Penumbra, and receives research grant support from National Institutes of Health (NINDS R25NS065743), Society of Vascular and Interventional Neurology, and Heitman Stroke Foundation. MDS is supported by the Heinz Family Foundation, Heitman Stroke Foundation, and NIA R21AG083559.